\documentclass[acmsmall]{acmart}

\usepackage{hyperref}
\usepackage{amsmath,amsfonts}
\usepackage{algorithmic}
\usepackage{graphicx}
\usepackage{textcomp}
\usepackage{xcolor}
\usepackage{makecell}
\usepackage{tcolorbox}
\usepackage{lipsum}
\usepackage{listings}
\usepackage{etoolbox}
\usepackage{multirow}
\usepackage{colortbl}
\usepackage{booktabs}
\usepackage{wrapfig}
\usepackage[framemethod=tikz]{mdframed}
\usepackage{adjustbox}
\usepackage{mdframed}

\AtBeginDocument{%
  }

\begin{document}

\title[Large Language Models in Fault Localisation]{Large Language Models in Fault Localisation}

\author{Yonghao Wu}
\email{appmlk@outlook.com}
\affiliation{%
  \institution{Beijing University of Chemical Technology}
  \city{Beijing}
  \country{China}
}

\author{Zheng Li}
\email{lizheng@mail.buct.edu.cn}
\affiliation{%
  \institution{Beijing University of Chemical Technology}
  \country{China}
}

\author{Jie M. Zhang}
\email{jie.zhang@kcl.ac.uk}
\affiliation{%
  \institution{University College London}
  \city{London}
  \country{UK}
}

\author{Mike Papadakis}
\email{michail.papadakis@uni.lu}
\affiliation{%
  \institution{University of Luxembourg}
  \country{Luxembourg}
}

\author{Mark Harman}
\email{mark.harman@cs.ucl.ac.uk}
\affiliation{%
  \institution{University College London}
  \city{London}
  \country{UK}
}

\author{Yong Liu}
\email{lyong@mail.buct.edu.cn}
\authornotemark[1]
\affiliation{%
  \institution{Beijing University of Chemical Technology}
  \city{Beijing}
  \country{China}
}


\begin{abstract}

Large Language Models (LLMs) have shown promise in multiple software engineering tasks including code generation, program repair, code summarisation, and test generation.
Fault localisation is instrumental in enabling automated debugging and repair of programs and was prominently featured as a highlight during the launch event of ChatGPT-4.
Nevertheless, the performance of LLMs compared to state-of-the-art methods, as well as the impact of prompt design and context length on their efficacy, remains unclear.
To fill this gap, this paper presents an in-depth investigation into the capability of ChatGPT-3.5 and ChatGPT-4, the two state-of-the-art LLMs, on fault localisation.
Using the widely-adopted large-scale Defects4J dataset,
we compare the two LLMs with the existing fault localisation techniques.
We also investigate the consistency of LLMs in fault localisation, as well as how prompt engineering and the length of code context affect the fault localisation effectiveness.

Our findings demonstrate that  within function-level context, ChatGPT-4 outperforms all the existing fault localisation methods.
Additional error logs can further improve ChatGPT models' localisation accuracy and consistency,
with an average 46.9\% higher accuracy over the state-of-the-art baseline SmartFL on the Defects4J dataset in terms of $TOP$-$1$ metric.
However, when the code context of the Defects4J dataset expands to the class-level, ChatGPT-4's performance suffers a significant drop, with 49.9\% lower accuracy than SmartFL under $TOP$-$1$ metric.
These observations indicate that although ChatGPT can effectively localise faults under specific conditions, limitations are evident.
Further research is needed to fully harness the potential of LLMs like ChatGPT for practical fault localisation applications.
\end{abstract}

\ccsdesc[500]{Software and its engineering~Software testing and debugging}

\keywords{Large Language Model, Fault Localisation, ChatGPT, Empirical Study}


\maketitle

\section{Introduction}

Among all the software engineering activities, fault localisation is notable as an important phase in the debugging process of software systems, serving as a prerequisite for developers to fix program errors accurately~\cite{li2022fault} and also automatic program repair~\cite{assiri2017fault}.
Efficient and accurate fault localisation techniques can significantly improve the efficiency of software repair and maintenance~\cite{khan2021alloyfl,wu2022theoretical}.
Current fault localisation methods employ a variety of techniques, including statistical analysis~\cite{kim2021ahead}, coverage analysis~\cite{wu2020fatoc,papadakis2015metallaxis}, and machine learning algorithms~\cite{weimer2009automatically,ni2022best}.
However, a prevalent limitation of current fault localisation techniques is their adaptability, which is often restricted by specific programming languages, the quality of test cases, or code structures, consequently diminishing their effectiveness~\cite{7390282}.

Large Language Models (LLMs), exemplified by ChatGPT, have attracted substantial interest within the computational and data science communities due to their wide-ranging applications and powerful performance~\cite{ray2023chatgpt,feng2023investigating}.
Their remarkable effectiveness in understanding natural language~\cite{tunstall2022natural} and generating meaningful content~\cite{chakraborty2022natgen,fried2022incoder} has spurred interest across disciplines including software engineering, where they have shown promises particularly on code generation~\cite{tian2023chatgpt,dong2023self}, program repair~\cite{xie2023impact,cao2023study,sobania2023analysis,fse2023copilot}, code summarisation~\cite{hu2020deep,sun2023automatic,sridhara2023chatgpt}, and test generation~\cite{xie2023chatunitest,schafer2023adaptive,liu2023your}.

LLMs also have the potential in aiding fault localisation.
In the ChatGPT-4 launch event, OpenAI demonstrated the model’s capability to localise and rectify faults, given knowledge of a program’s code and error log.
This demonstration unveiled the potential and applicability of ChatGPT-4 in fault localisation.
Nevertheless, it remains unclear how the performance of LLMs compares to the state-of-the-art, and how performances are influenced by prompt design and context length.


To fill this gap, in this paper, we comprehensively study the performance of ChatGPT in fault localisation.
The findings reveal its current advantages, shortcomings, and provide insights for future research.
Our study aims to answer the following research questions.

\begin{itemize}

\item \textbf{RQ1. How does ChatGPT perform in fault localisation?}

\begin{itemize}

\item \textbf{RQ1.1. How does the accuracy of ChatGPT's fault localisation compare to state-of-the-art methods?}
This RQ aims to examine the capability of ChatGPT-3.5 and ChatGPT-4 in localising faults within programs from the Defects4J dataset.
Using the $TOP$-$N$ metric, a commonly accepted standard of evaluation, we can quantify the number of accurately localised faulty functions.

\item \textbf{RQ1.2. How do ChatGPT's fault localisation results overlap with state-of-the-art methods?}
This RQ aims to conduct a more detailed analysis, building on the general performance comparison offered by RQ1.1.
Specifically, we investigate the number of programs where ChatGPT performs uniquely or is missing compared to other baselines.
This comparative analysis can further reveal the situations and reasons existing methods surpass ChatGPT.

\item \textbf{RQ1.3. How consistent is ChatGPT's performance in fault localisation across repeated experiments?}
To assess ChatGPT's reliability in fault localisation, we conduct five repeated experiments in RQ1.1.
This RQ focuses on identifying potential inconsistencies in ChatGPT's performance and understanding how consistency manifests over experiments.

\end{itemize}

\item \textbf{RQ2. How is ChatGPT's fault localisation performance affected by variations in prompt design?}
This RQ examines the experimental impact of omitting individual components from a standard prompt on ChatGPT's fault localisation ability.
Through evaluating the model's performance with various optional components excluded, we aim to identify which components are crucial for achieving optimal results.

\item \textbf{RQ3. How does the length of code context in prompts influence ChatGPT's fault localisation performance?} 
This RQ explores how expanding or narrowing the scope of code context in prompts affects ChatGPT's performance in localising faults within large codebases.
We evaluate ChatGPT's accuracy when the context of the Defects4J dataset is expanded to class-level context versus narrowed to only statements surrounding the fault.
The results will reveal insights into ChatGPT's capabilities and limitations for practical fault localisation in varying scales of software systems.
\end{itemize}

In addition, to alleviate the potential overfitting threat that Defects4J being part of ChatGPT's training data, we have collected a more recent dataset named ``StuDefects'', in which all the programs were written in and after 2022, hence are outside the timeframe of ChatGPT's data acquisition period. An extended analysis has been carried out on this new dataset to validate our conclusion.

The contributions of this work are summarised as follows:

\begin{itemize}
\item We conduct a comprehensive empirical study on the large-scale open-source program Defects4J, to evaluate the potential of LLM models for fault localisation research.
Our results demonstrate that, with function-level context, ChatGPT-4, when augmented with test cases and error logs, achieves a mean value of 23.13 in terms of $TOP$-$1$ evaluation metric, outperforming the state-of-the-art baseline SmartFL by an average of 46.9\%.
However, when the code context is expanded from function level to class level, the effectiveness of ChatGPT-4 deteriorates markedly, performing 49.9\% worse than SmartFL.

\item We assess the consistency of ChatGPT for fault localisation.
Our findings suggest that incorporating dynamic execution information into ChatGPT enables more consistent fault localisation results, reducing the average variance in $TOP$-$1$ metric to 12.00\% for ChatGPT-4, the lowest among ChatGPT related methods.

\item We evaluate the influence of each component in the prompt on the fault localisation capabilities of ChatGPT. 
Our results indicate that excluding any component leads to a decline in the performance of ChatGPT-4, as measured by $TOP$-$1$ metric.
Among them, excluding the error log causes the most significant impact, resulting in a 25.6\% reduction in accuracy.

\item We alleviate the concern of ChatGPT's potential overfitting to Defects4J by 
introducing a novel, more recent dataset named ``StuDefects''. Our extended analysis on this new dataset
yields consistent observations with those from Defects4J (52.9\% superiority over the state-of-the-art in terms of $TOP$-$1$).
\end{itemize}

The data and code supporting the findings of this study, including the original responses from ChatGPT, are publicly available on GitHub~\cite{thisweb}.

The remainder of this paper is organised as follows. 
Section~\ref{Design} outlines the experimental design, including descriptions of the dataset, baseline models, ChatGPT setup and evaluation metrics used in this study.
Section~\ref{Result} delves into a detailed analysis of the results for each RQ.
Section~\ref{Extended Analysis} provides an extended analysis of the generalisability of ChatGPT on the new dataset we collected.
Section~\ref{RelatedWork} reviews related prior work. 
Section~\ref{Discussion} discusses threats to validity and implications for applying LLMs to fault localisation.
Section~\ref{conclusion} summarises our study and discusses potential future work.

\section{Experimental Design}
\label{Design}

\subsection{Dataset}
\label{Dataset}


Defects4J~\cite{Defects4Jweb} is a widely used fault localisation dataset containing real-world open-source programs with real bugs that occurred in their history versions and corresponding test cases ~\cite{just2014defects4j}.

\begin{table}[htp]
 \scriptsize
\caption{Statistic of the Defect4J Dataset}
\label{dataset}
\centering
\renewcommand\arraystretch{1}
\begin{tabular}{r|c|c|c}
\hline 
Program & \makecell[c]{Number of\\Versions} &\makecell[c]{ Number of\\Faulty Functions} & \makecell[c]{Average Length\\of Code}\tabularnewline
\hline 
Chart & 17 & 21 & 62,395\tabularnewline
Lang & 35 & 52 & 13,441\tabularnewline
Math & 66 & 89 & 36,226\tabularnewline
Mockito & 10 & 33 & 6,722\tabularnewline
Time & 15 & 24 & 20,357\tabularnewline
Closure & 52 & 61 & 64,538\tabularnewline
\hline 
\textbf{Sum} & 195 & 280 & -\tabularnewline
\hline 
\end{tabular}
\end{table}

In this study, we use six large-scale open-source Java projects from Defects4J for experimental analysis, summarized in Table~\ref{dataset}.
These programs are  JFreeChart (Chart), Google Closure compiler (Closure), Apache Commons-Lang (Lang), Apache Commons-Math (Math), Mockito Framework (Mockito), and Joda-Time (Time).

The first column of Table~\ref{dataset} lists these six programs.
The second, third, and fourth columns provide the corresponding number of available versions, the number of faulty functions, and the average lines of code for each program, respectively.

These projects contain multiple program versions, where versions that fail to compile or have segmentation errors are excluded. 
We take the intersection of program versions supported by each baseline as the final dataset, which contains a total of 195 program versions.
Additionally, these projects vary in scale, complexity, and code length.
We use the Defects4J dataset as it is widely used in research~\cite{Grace2021,zou2019empirical,zakari2020multiple}, making our results comparable with much existing literature, providing a dependable evaluation of fault localisation techniques.


\subsection{Baselines}

\subsubsection{Spectrum-Based Fault Localisation (SBFL)}

SBFL is a debugging technique that leverages execution traces of passed and failed test cases to estimate the likelihood of program statements being faulty.
The core intuition is that faulty statements are more frequently covered by failed tests compared to passed tests.
SBFL techniques quantify this intuition by computing the suspiciousness value of each program statement.
Common suspiciousness formulas of SBFL proposed in literature include $Jaccard$~\cite{Jaccard}, $Tarantula$~\cite{1007991}, $Ochiai$~\cite{4041886}, $OP2$~\cite{op2}, and $Dstar$~\cite{wong2013dstar}.
Then, SBFL ranks program statements based on their suspiciousness values.
Developers can then inspect the ranked list, starting from the most suspicious statements to identify potential bug locations.

In this study, we adopt $Ochiai$ and $Dstar$ formulas of SBFL to calculate the suspiciousness value of program statements, as these formulas have demonstrated the best performance in existing research~\cite{Pearson2017Evaluating,zou2019empirical}.
The formulas of $Ochiai$ and $Dstar$ in SBFL are:

\begin{minipage}{0.5\linewidth}
\small
\[
Ochiai(statement)=\frac{e_{f}}{\sqrt{(e_{f}+n_{f})\times(e_{f}+e_{p})}}
\]
\end{minipage}%
\begin{minipage}{0.5\linewidth}
\small
\[
Dstar(statement)=\frac{e_{f}^{*}}{n_{f}+e_{p}}
\]
\end{minipage}%

Where \textit{Ochiai(statement)} or \textit{Dstar(statement)} signify the suspiciousness values for a given program statement.
The variables $e_{p}$ and $e_{f}$ denote the counts of passed and failed test cases covering this statement.
Conversely, $n_{f}$ indicates the count of failed test cases not covering the program statement.
For the $Dstar$ formula, the exponent ``*'' is set to 3, as recommended by Wong et al.~\cite{wong2013dstar}.

A major advantage of SBFL is its lightweight instrumentation to collect coverage traces without needing program semantics.
One limitation is that SBFL relies on the quality of the test suite - its effectiveness is constrained by how well the test cases execute the code and reveal faults.

\subsubsection{Mutation-Based Fault Localisation (MBFL)}

MBFL localises potential bugs by leveraging program mutants~\cite{moon2014ask,papadakis2015metallaxis}.
Traditional MBFL automatically seeds bugs by applying mutation operators such as replacing operators.
The mutated programs (mutants) are executed against the test suite to identify mutants killed by test cases.
By comparing test outputs between the original and mutants, MBFL allocates program statements frequently impacted by killed mutants with higher suspiciousness values.
These statements receive higher mutation scores indicative of their correlation with bugs.
A salient strength of MBFL is its ability to achieve high fault localisation accuracy on traditional datasets such as Software-artifact Infrastructure Repository (SIR)~\cite{papadakis2015metallaxis}.

In this study, we refer to the experimental results of existing research and choose Metallaxis, which performs best in MBFL to calculate the suspiciousness of a statement~\cite{zou2019empirical,jia2020smfl,Pearson2017Evaluating}.
For each specific program statement, MBFL generates a set of mutants $M(statement) = \langle m_1, m_2, \ldots \rangle $ where $m_i \in M(statement)$.
MBFL subsequently evaluates the suspiciousness of each mutant using the Metallaxis formula and attributes the highest such value to the original program statement, thereby representing its likelihood of being faulty.
The formulas of $Ochiai$ and $Dstar$ in MBFL are:

\begin{minipage}{0.5\linewidth}
\scriptsize
\[
Ochiai(m_{i})=\frac{failed(m_{i})}{\sqrt{totalfailed\times(failed(m_{i})+passed(m_{i}))}}
\]
\end{minipage}
\begin{minipage}{0.5\linewidth}
\scriptsize
\[
Dstar(m_{i})=\frac{failed(m_{i})^{*}}{(totalfailed-failed(m_{i}))+passed(m_{i})}
\]
\end{minipage}

\vspace{0.2cm}
Where $failed(m_i)$ represents the count of test cases that failed on the original program and exhibit altered output when run on a mutant $m_i$.
$passed(m_i)$ represents the count of test cases that passed on the original program but exhibit different output when run on the mutant $m_i$.
$totalfailed$ denotes the aggregate count of test cases that fail when executed on the original program.

However, the process of generating and executing a multitude of mutants entails substantial computational overhead~\cite{liu2018optimal}.
Moreover, the effectiveness of MBFL is closely tied to the killing power of both the test suite and the generated mutants.
While MBFL achieves promising results on datasets with artificially seeded faults, recent studies using real-world defects indicate that MBFL underperforms SBFL on Defects4J dataset.
For example, research by Zou et al.~\cite{zou2019empirical} demonstrates that in the Defects4J dataset, SBFL outperforms MBFL across $TOP$-$3$, $TOP$-$5$, and $TOP$-$10$ evaluation metrics.
Specifically, the $Ochiai$ formula within SBFL is able to localise 156 faults when assessed by $TOP$-$10$ metric, while the most effective technique within MBFL localises 129 faults.
Existing research analyzes the limitations of traditional MBFL.
These methods rely on a limited set of manually-designed operators that focus on simple syntactic changes, failing to adequately represent real faults due to their lack of semantic depth~\cite{tian2022learning}.

\subsubsection{Semantics-based probabilistic Fault Localisation (SmartFL)}

Zeng et al.~\cite{zeng2022fault} proposed SmartFL, a novel statement-level fault localisation approach that models the likelihood of faulty program statements using probabilistic inference.
It combines static analysis and dynamic execution traces to estimate the correctness of program values without capturing full semantics. SmartFL achieves a balance between effectiveness and scalability.
Their paper shows that SmartFL outperforms existing spectrum-based, mutation-based, and semantics-based methods on Defects4J dataset, achieving 21\% $TOP$-$1$ statement accuracy, 5\% higher than the next best method.

\subsection{ChatGPT Setup}

In this study, we explore four fault localisation methods using different versions and prompt strategies of ChatGPT, named ChatGPT-3.5 (Origin), ChatGPT-3.5 (Log), ChatGPT-4 (Origin), and ChatGPT-4 (Log).
ChatGPT-3.5 and ChatGPT-4 denote two different versions of ChatGPT.
``Origin'' means that the information provided with ChatGPT in the prompt is merely the target function along with basic instructions,
while ``Log'' means that we further provide ChatGPT with a failed unit test case and the corresponding error log in a follow-up dialogue. 
In the following, we introduce the details for our setup  as well as specific examples of our designed prompts.



\subsubsection*{\textbf{Access to ChatGPT}}
For each version of ChatGPT, to get optimal performance, we conducted a preliminary study to compare the performance accessed via API with that via web interfaces using the Chart program from Defects4J 
Our results reveal that ChatGPT-3.5 has comparable results between the web interface and the API (e.g., 4.6 v.s. 4.8), while the web interface for ChatGPT-4 has better performance than API (12.4 v.s. 8 with log information). 
Therefore, we adopt the API for all ChatGPT-3.5 experiments for efficiency and choose the  web interface for ChatGPT-4 due to its superior performance.
For the web interface of ChatGPT-4, we have double-checked and guaranteed that the model behind it remains identical throughout our experimental procedure.


\subsubsection*{\textbf{Code Context}}

Existing LLMs often have limitations in the length of tokens they can accept in the prompt.
For example, ChatGPT-3.5 and ChatGPT-4 have a token limit of 4096 and 8192, respectively.
As a result, this prevents the application of LLMs on large-scale codebases such as those in Defects4J, which may contain tens of thousands of lines.

We address this issue, in line with state-of-the-art LLM-based research in software engineering~\cite{prenner2022can,xia2022practical}, by extracting the function containing faulty statements in the program under test.
This function provides the context in the prompt, effectively narrowing the scope of the code context for ChatGPT to localising faulty statements within the target function.
In addition, in order to ensure a fair comparison, in the subsequent experimental analyses, the fault localisation scope across all baselines is kept consistent with that of ChatGPT.

Furthermore, in RQ3, we perform comparative experiments on the Defects4J dataset to analyse the impact of varying the length of code context in prompts on ChatGPT's performance.

\subsubsection*{\textbf{Design of Prompt}}
For ChatGPT-3.5/4 (Origin),
the prompt we privide includes the complete target function and basic instructions to ChatGPT.
We call this prompt $Prompt_1$.
The basic instructions in $Prompt_1$ request ChatGPT to present potential faulty locations sorted in descending order of suspicion.
Additionally, an explanation is required to explain how each location is pinpointed as potentially faulty, 
a choice inspired by Gao et al.~\cite{gao2023self} where they found that 
self-explanatory prompts significantly enhance LLMs' comprehension in complex dialogues.


To illustrate, an example of $Prompt_1$ is provided below.

\definecolor{texgray}{RGB}{244, 245, 246}
\definecolor{texrightgray}{RGB}{99, 99, 99}

\definecolor{green}{RGB}{214, 230, 189}
\definecolor{leftgreen}{RGB}{19, 138, 7}
\definecolor{gray}{RGB}{200,200,200}

\newmdenv[
  leftline=false,
  rightline=true,
  topline=false,
  bottomline=false,
  backgroundcolor=texgray, 
  linecolor=texrightgray, 
  linewidth=2pt,
  skipabove=5pt
]{myrightline}

\lstnewenvironment{code}[1][]
{\lstset{
  language=Java,
  basicstyle=\scriptsize\ttfamily,
  breaklines=true,
  numbers=left,
  numbersep=5pt,
  xleftmargin=10pt,
  lineskip=-10pt, 
}}
{}

\lstdefinelanguage{json}{
    basicstyle=\scriptsize\normalfont\ttfamily,
    numberstyle= \scriptsize,
    stepnumber=1,
    numbersep=8pt,
    showstringspaces=false,
    breaklines=true,
    string=[s]{"}{"},
    comment=[l]{//},
    morecomment=[s]{/*}{*/},
    emphstyle=\bfseries\color{blue},
    tabsize=2,
    lineskip=-10pt, 
}

\begin{myrightline}
\footnotesize
\textbf{Example of $Prompt_{1}$:}

Please analyse the following code snippet for potential bugs. Return the results in JSON format, consisting of a single JSON object with two fields: ``intentOfThisFunction'' (describing the intended purpose of the function), and ``faultLocalization'' (an array of JSON objects). The ``faultLocalization'' array should contain up to five JSON objects, each with three fields: ``lineNumber'' (indicating the line number of the suspicious code), ``codeContent'' (showing the actual code), and ``reason'' (explaining why this location is identified as potentially faulty). Note: The codes in the ``faultLocalization'' array should be listed in descending order of suspicion.
\begin{code}
public static String determineTriangleType(double a, double b, double c) {
    if (a+b<c || a+c<b || b+c<a) {
        return "Invalid triangle";
    }
    if (a==b && b==c) {
        return "Equilateral triangle";
    } else if (a==b || b==c || a==c) {
        return "Isosceles triangle";
    } else {
        return "Scalene triangle";
    }
}
\end{code}
\end{myrightline}

An answer from ChatGPT-4 to $Prompt_1$ for a function that determines the type of triangle based on side lengths is provided below.

\newmdenv[
  leftline=true,
  rightline=false,
  topline=false,
  bottomline=false,
  backgroundcolor=green!20, 
  linecolor=leftgreen, 
  linewidth=2pt,
  skipabove=5pt
]{myleftline}

\begin{myleftline}
\footnotesize
\textbf{Answer to $Prompt_{1}$:}\\
\includegraphics[width=1\linewidth]{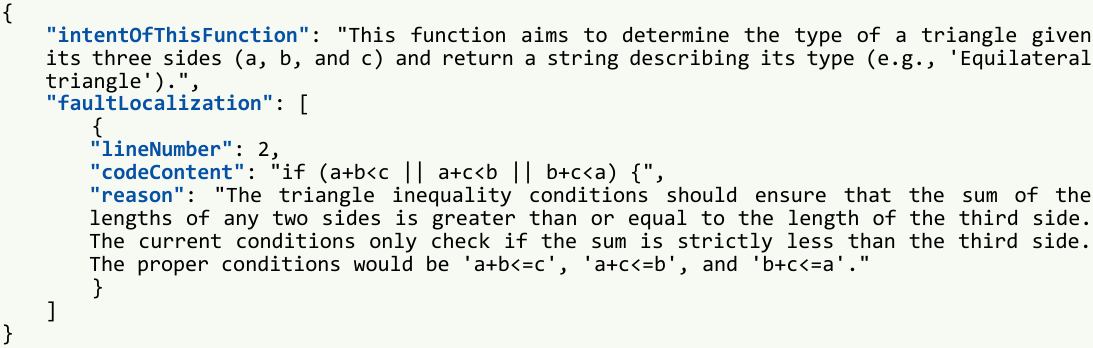}
\end{myleftline}

In this answer, ChatGPT identifies line 2 as a potential fault, since the current ``<'' condition cannot handle the invalid triangle case when the sum of the lengths of the two sides equals the third side.
This example demonstrates ChatGPT's capacity to provide reasoned fault localisation by analysing code logic and limitations.

For ChatGPT-3.5/4 (log), we aim to further reveal ChatGPT's fault localisation capability with another prompt, called  $Prompt_2$, that contains a corresponding failed unit test case and the error log derived from the test execution results.
$Prompt_2$ is provided to ChatGPT after the first interaction with $Prompt_1$ to let ChatGPT re-examine and refine its initial suspected faulty locations.
An example of $Prompt_2$ is illustrated below.

\begin{myrightline}
\footnotesize
\textbf{Example of $Prompt_{2}$:}

I have received an error message and a unit test case related to the code snippet I provided in the first prompt. The error message is:
\begin{lstlisting}[language=json]
org.junit.ComparisonFailure: 
Expected :Invalid triangle
Actual   :Isosceles triangle
	at org.junit.Assert.assertEquals(Assert.java:115)
	at org.junit.Assert.assertEquals(Assert.java:144)
	at com.example.TriangleTest.testDetermineTriangleType(TriangleTest.java:6)
	...
\end{lstlisting}
Additionally, here is the unit test case:
\begin{code}
import static org.junit.Assert.assertEquals;
import org.junit.Test;
public class TriangleTest {
    @Test
    public void testDetermineTriangleType() {
        assertEquals("Invalid triangle", determineTriangleType(-1.0,-1.0,1.0));
    }
}
\end{code}

Please analyse the code snippet from the first prompt, along with the provided error message and unit test case. Update and return the JSON object consisting of ``intentOfThisFunction'' (describing the intended purpose of the function), and ``faultLocalization'' (an array of JSON objects). The ``faultLocalization'' array should contain up to five JSON objects, each with three fields: ``lineNumber'' (indicating the line number of the suspicious code), ``codeContent'' (showing the actual code), and ``reason'' (explaining why this location is identified as potentially buggy). Note: The codes in the ``faultLocalization'' array should be listed in descending order of suspicion, and the analysis should focus exclusively on the code snippet from the first prompt and not the unit test case.

\end{myrightline}

\vspace{0.2cm}

In RQ2, we conduct an ablation study to check how different components in the prompts affect fault localisation.

\subsubsection*{\textbf{Collection of Results}}
For convenient management of experimental results, we instruct ChatGPT to provide fault localisation outputs in the universally recognised JSON format.
The response is expected to adhere to the format and content specified in the prompt requirements.
In case of any deviations, we continue to resubmit the prompt until the output aligns with the given specifications.
Furthermore, some target functions in Defects4J may avoid execution by test cases or not trigger exceptions, failing to yield error log information.
In such cases, the fault localisation results of ChatGPT-3.5/4 (Log) equal to those of ChatGPT-3.5/4 (Origin).


After getting the JSON result given by ChatGPT, we parse the JSON string and follow the instructions of $Prompt_1$ to take the first five code statements in the JSON string as the fault localisation result of ChatGPT.
Finally, we can use this result to evaluate the performance of ChatGPT in terms of $TOP$-$1$ to $TOP$-$5$ metrics ($TOP$-$N$ metric will be described in Section~\ref{TOP-N}).

\subsubsection*{\textbf{Mitigation of Randomness}}
Due to ChatGPT's inherent randomness, we adopt a rigorous method with repeated experiments for each target function. 
Specifically, all ChatGPT experiments in this paper are performed five times independently.
We then average the $TOP$-$N$ metric as the final fault localisation outcome.
This ensures reliable results by minimizing ChatGPT's random bias.

\subsection{Measurement of Fault Localisation Performance}

\subsubsection{TOP-N}
\label{TOP-N}
We use the widely adopted $TOP$-$N$ metric to measure the accuracy of fault localisation techniques~\cite{zeng2022fault,Grace2021,wu2020fatoc}.
$TOP$-$N$ metric counts the number of faulty functions that contain faulty statements with a suspiciousness rank less than or equal to $N$ ~\cite{li2017transforming}.
In other words, $TOP$-$N$ refers to the number of faulty functions that developers can localise the exact position of faulty statements by checking the top $N$ statements.
Therefore, higher $TOP$-$N$ values indicate more effective fault localisation.
Prior studies indicate that 70\% of developers and testers focus only on the top five program statements in fault localisation ranking lists ~\cite{10.1145/2931037.2931051}.
Thus, we focus our evaluation metrics from $TOP$-$1$ to $TOP$-$5$.

It is worth noting that the baselines in this paper (i.e., SBFL, MBFL, and SmartFL) may assign the same suspiciousness score to correct statements as to faulty ones~\cite{steimann2013threats,7390282}.
For these circumstances, we use the average ranking strategy, which is commonly employed in existing studies~\cite{7390282,Pearson2017Evaluating}. That is, we use the average ranks of all statements with the same suspiciousness score as the final ranks of faulty statements.

\subsubsection{Wilcoxon Signed-Rank Test}
\label{subsec:Wilcoxon-Signed-Rank-Test}

In this study, we use the Wilcoxon signed-rank test, a widely adopted non-parametric statistical analysis approach in fault localisation~\cite{kim2019vfl,DeepFL2019,gao2018mseer},
 to determine the confidence level concerning the varied effectiveness of different fault localisation methods.
The Wilcoxon signed-rank test offers a robust evaluation
that is specifically useful when the normal distribution of data cannot be assumed
~\cite{ott2015introduction}.
It assesses if two related samples come from identical populations by comparing ranks of paired differences, and is ess sensitive to outliers than mean comparisons.

\section{Results Analysis}
\label{Result}

\subsection{RQ1: Effectiveness of ChatGPT in Fault Localisation}

\noindent
\textbf{RQ1.1: Comparison with Baselines:} 
In this section, we comprehensively compare the efficacy of different fault localisation techniques.
Table~\ref{generalresults} presents the results on the six Defects4J programs, as measured by $TOP$-$N$ metric.
The performance metrics for ChatGPT-3.5 and ChatGPT-4 are the averages of five independent repeated experiments, thus ensuring the statistical significance of our results.
Additionally, in the table, the best-performing method in each row, as indicated by the largest $TOP$-$N$ value, is highlighted in green.
The state-of-the-art baseline SmartFL was not implemented in two of the programs (Mockito and Closure) in its proposed paper, and we use ``-'' as a substitute in the corresponding situations.

\begin{table}[htp]
 \tiny
\caption{RQ1.1: Fault Localisation Effectiveness on Defects4J }
\label{generalresults}
\centering
\renewcommand\arraystretch{1}
\begin{tabular}{c|c|cc|cc|c|cc|cc}
\hline 
\multirow{3}{*}{Program} & \multirow{3}{*}{$TOP$-$N$} & \multicolumn{9}{c}{Method}\tabularnewline
 &  & \multicolumn{2}{c|}{SBFL} & \multicolumn{2}{c|}{MBFL} & \multirow{2}{*}{SmartFL} & \multicolumn{2}{c|}{ChatGPT-3.5} & \multicolumn{2}{c}{ChatGPT-4}\tabularnewline
 &  & $Ochiai$ & $Dstar$ & $Ochiai$ & $Dstar$ &  & Origin & Log & Origin & Log\tabularnewline
\hline 
\multirow{5}{*}{Chart} & $TOP$-$1$ & 3 & 3 & 3 & 3 & 8 & 4.4 & 4.8 & 11 &\cellcolor[HTML]{d9ead3} 12.4\tabularnewline
 & $TOP$-$2$ & 8 & 8 & 11 & 11 & 9 & 6.2 & 8.8 & 14.4 & \cellcolor[HTML]{d9ead3} 15.2\tabularnewline
 & $TOP$-$3$ & 9 & 8 & 14 & 14 & 11 & 8.2 & 10.4 & 16.2 & \cellcolor[HTML]{d9ead3} 17\tabularnewline
 & $TOP$-$4$ & 11 & 12 & 14 & 14 & 12 & 8.8 & 10.8 & 16.4 & \cellcolor[HTML]{d9ead3} 17.2\tabularnewline
 & $TOP$-$5$ & 13 & 13 & 14 & 14 & 13 & 10 & 11.4 & 16.6 & \cellcolor[HTML]{d9ead3} 17.6\tabularnewline
\hline 
\multirow{5}{*}{Lang} & $TOP$-$1$ & 8 & 8 & 9 & 7 & 19 & 13.6 & 13.6 & 17.6 & \cellcolor[HTML]{d9ead3} 26.6\tabularnewline
 & $TOP$-$2$ & 17 & 16 & 26 & 25 & 29 & 20.2 & 20.2 & 21.4 & \cellcolor[HTML]{d9ead3} 31.4\tabularnewline
 & $TOP$-$3$ & 23 & 22 & 31 & 30 & 31 & 23.8 & 24.8 & 24.2 & \cellcolor[HTML]{d9ead3} 35\tabularnewline
 & $TOP$-$4$ & 27 & 27 & 35 & 34 & 34 & 26 & 26.6 & 26.6 & \cellcolor[HTML]{d9ead3} 36.8\tabularnewline
 & $TOP$-$5$ & 29 & 28 & 39 & 38 & 38 & 27 & 28.4 & 27.6 & \cellcolor[HTML]{d9ead3} 37.4\tabularnewline
\hline 
\multirow{5}{*}{Math} & $TOP$-$1$ & 24 & 20 & 25 & 25 & 32 & 26.8 & 30.8 & 28.4 & \cellcolor[HTML]{d9ead3} 41.8\tabularnewline
 & $TOP$-$2$ & 37 & 34 & 44 & 44 & 42 & 41.2 & 43.4 & 45.2 & \cellcolor[HTML]{d9ead3} 58.8\tabularnewline
 & $TOP$-$3$ & 44 & 39 & 53 & 53 & 51 & 46.2 & 48 & 54.2 & \cellcolor[HTML]{d9ead3} 64.4\tabularnewline
 & $TOP$-$4$ & 48 & 43 & 57 & 57 & 56 & 50.2 & 52.4 & 59 & \cellcolor[HTML]{d9ead3} 69.6\tabularnewline
 & $TOP$-$5$ & 57 & 51 & 65 & 65 & 62 & 54 & 56 & 61.2 & \cellcolor[HTML]{d9ead3} 72\tabularnewline
\hline 
\multirow{5}{*}{Mockito} & $TOP$-$1$ & 16 & 16 & 8 & 8 & - & 22.6 & 23.8 & 21.8 & \cellcolor[HTML]{d9ead3} 28.4\tabularnewline
 & $TOP$-$2$ & 27 & 26 & 25 & 24 & - & 27.2 & 27.8 & 24.2 & \cellcolor[HTML]{d9ead3} 29.2\tabularnewline
 & $TOP$-$3$ & 30 & 29 & 26 & 26 & - & 28.6 & 29.2 & 26.8 & \cellcolor[HTML]{d9ead3} 30.2\tabularnewline
 & $TOP$-$4$ & 30 & 29 & 27 & 27 & - & 29.2 & 29.8 & 28.2 & \cellcolor[HTML]{d9ead3} 30.4\tabularnewline
 & $TOP$-$5$ & \cellcolor[HTML]{d9ead3} 31 & 30 & 28 & 28 & - & 29.6 & 30 & 28.6 &  30.4\tabularnewline
\hline 
\multirow{5}{*}{Time} & $TOP$-$1$ & 5 & 5 & 5 & 5 & 4 & 3.8 & 5.4 & 4.2 & \cellcolor[HTML]{d9ead3} 7.6\tabularnewline
 & $TOP$-$2$ & 5 & 7 & 6 & 6 & 6 & 6.8 & \cellcolor[HTML]{d9ead3} 8.6 & 5.6 & \cellcolor[HTML]{d9ead3} 8.6\tabularnewline
 & $TOP$-$3$ & 10 & 10 & 9 & 9 & 7 & 8.2 & \cellcolor[HTML]{d9ead3} 10.6 & 7.2 & 10.2\tabularnewline
 & $TOP$-$4$ &  12 & 12 & 11 & 11 & 11 & 10 & \cellcolor[HTML]{d9ead3} 12.4 & 8.6 & 12\tabularnewline
 & $TOP$-$5$ & \cellcolor[HTML]{d9ead3} 18 & \cellcolor[HTML]{d9ead3} 18 & 17 & 17 & 16 & 11.8 & 14.2 & 13.2 & 16.4\tabularnewline
\hline 
\multirow{5}{*}{Closure} & $TOP$-$1$ & 17 & 14 & 1 & 1 & - & 10.4 & 9.6 & 12.6 & \cellcolor[HTML]{d9ead3} 22\tabularnewline
 & $TOP$-$2$ & 22 & 18 & 6 & 6 & - & 17 & 17.6 & 18.8 & \cellcolor[HTML]{d9ead3} 28.6\tabularnewline
 & $TOP$-$3$ & 25 & 22 & 10 & 10 & - & 21 & 23 & 23.8 & \cellcolor[HTML]{d9ead3} 32.6\tabularnewline
 & $TOP$-$4$ & 34 & 30 & 11 & 11 & - & 24.4 & 26 & 26.8 & \cellcolor[HTML]{d9ead3} 34.4\tabularnewline
 & $TOP$-$5$ & \cellcolor[HTML]{d9ead3} 38 & 33 & 14 & 14 & - & 27 & 27.2 & 29 &  35.8\tabularnewline
\hline
\hline
\multirow{5}{*}{\textbf{Average}} & $TOP$-$1$ & 12.17  & 11.00  & 8.50  & 8.17  & 15.75  & 13.60  & 14.67  & 15.93  & \cellcolor[HTML]{d9ead3} 23.13 \tabularnewline
 & $TOP$-$2$ & 19.33  & 18.17  & 19.67  & 19.33  & 21.50  & 19.77  & 21.07  & 21.60  & \cellcolor[HTML]{d9ead3} 28.63 \tabularnewline
 & $TOP$-$3$ & 23.50  & 21.67  & 23.83  & 23.67  & 25.00  & 22.67  & 24.33  & 25.40  & \cellcolor[HTML]{d9ead3} 31.57 \tabularnewline
 & $TOP$-$4$ & 27.00  & 25.50  & 25.83  & 25.67  & 28.25  & 24.77  & 26.33  & 27.60  & \cellcolor[HTML]{d9ead3} 33.40 \tabularnewline
 & $TOP$-$5$ & 31.00  & 28.83  & 29.50  & 29.33  & 32.25  & 26.57  & 27.87  & 29.37  & \cellcolor[HTML]{d9ead3} 34.93 \tabularnewline
\hline 
\end{tabular}
\end{table}

The results presented in Table \ref{generalresults} show that ChatGPT-4 (Log) yield the most promising results, highlighting the significant advancements made in developing enhanced fault localisation techniques.
ChatGPT-3.5 (Origin) and ChatGPT-3.5 (Log) achieve optimal performance in limited cases.
Meanwhile, SmartFL sets a higher contemporary benchmark for overall efficacy over traditional fault localisation techniques SBFL and MBFL.

\noindent
\vspace{0.2cm}
\begin{mdframed}[backgroundcolor=texgray, linewidth=0.5pt]
\textbf{Answer to RQ1.1:}
ChatGPT-4 (Log), i.e., providing ChatGPT-4 with the log information of the failed tests, yields the best effectiveness in fault localisation on Defects4J. 
The mean value of $TOP$-$1$ for ChatGPT-4 (Log) is 23.13, outperforming the state-of-the-art baseline SmartFL by 46.9\%.
\end{mdframed}

\noindent
\textbf{RQ1.2: Overlap of Different Methods:}
In this section, we analyze the overlap between ChatGPT-4 (Log), the best-performing way of using ChatGPT in fault localisation, and other methods in fault localisation, with the total 186 faults.
Table~\ref{Overlap Result} shows the results, where
Column ``Unique'' shows the number and ratio of faults where ChatGPT-4 (Log) can locate while other methods cannot.
Take the first cell as an example, the value ``34'' indicates that
there are 34 (18.28\%) bugs ChatGPT-4 (Log) can locate but ChatGPT-4 (Origin) cannot in terms of $Top$-$1$.
Column ``Overlapping'' shows the number and ratio of faults where both ChatGPT-4 (Log) and the given baseline method are able to localise.
Column ``Missing'' shows the number and ratio of faults that only the baseline method can locate, while ChatGPT-4 (Log) cannot. 
Experiments in this section do not include programs not implemented in SmartFL (i.e., Mockito and Closure).
Consequently, there are a total of 186 remaining faulty versions.

\begin{wraptable}{l}{9cm}
\scriptsize
\caption{RQ1.2: Overlap Analysis of ChatGPT-4 (Log) and Other Methods. }
\label{Overlap Result}
\centering
\renewcommand\arraystretch{1}
\begin{tabular}{l|c|ccc}
\hline 
\makecell[l]{Alternative methods} & Metric & Unique & Overlapping & Missing\tabularnewline
\hline 
\multirow{2}{*}{ChatGPT-4 (Origin)} & $TOP-1$ & 34 (18.28\%) & 145 (77.96\%) & 7 (3.76\%)\tabularnewline
 & $TOP-5$ & 28 (15.05\%) & 155 (83.33\%) & 3 (1.61\%)\tabularnewline
\hline 
\multirow{2}{*}{ChatGPT-3.5 (Origin)} & $TOP-1$ & 53 (28.49\%) & 120 (64.52\%) & 13 (6.99\%)\tabularnewline
 & $TOP-5$ & 55 (29.57\%) & 117 (62.9\%) & 14 (7.53\%)\tabularnewline
\hline 
\multirow{2}{*}{ChatGPT-3.5 (Log)} & $TOP-1$ & 48 (25.81\%) & 123 (66.13\%) & 14 (7.53\%)\tabularnewline
 & $TOP-5$ & 50 (26.88\%) & 120 (64.52\%) & 16 (8.6\%)\tabularnewline
\hline 
\multirow{2}{*}{SBFL ($Ochiai$)} & $TOP-1$ & 61 (32.8\%) & 112 (60.22\%) & 13 (6.99\%)\tabularnewline
 & $TOP-5$ & 48 (25.81\%) & 116 (62.37\%) & 22 (11.83\%)\tabularnewline
\hline 
\multirow{2}{*}{MBFL ($Ochiai$)} & $TOP-1$ & 59 (31.72\%) & 114 (61.29\%) & 13 (6.99\%)\tabularnewline
 & $TOP-5$ & 29 (15.59\%) & 136 (73.12\%) & 21 (11.29\%)\tabularnewline
\hline 
\multirow{2}{*}{SmartFL} & $TOP-1$ & 44 (23.66\%) & 124 (66.67\%) & 18 (9.68\%)\tabularnewline
 & $TOP-5$ & 37 (19.89\%) & 126 (67.74\%) & 23 (12.37\%)\tabularnewline
\hline 
\end{tabular}
\end{wraptable}


As observed from Table~\ref{Overlap Result}, the values of ``Unique'' are all larger than the values of ``Missing'', indicating that ChatGPT-4 (Log) successfully localises more unique faults than the other methods. 
Further, ChatGPT-4 (Log) and ChatGPT-4 (Origin) have the highest ``Overlapping'' values, signifying consistent performance in more cases.
Additionally, among all baselines, ChatGPT-4 (Log) has the largest ``Missing'' value when compared to SmartFL.
This suggests that, compared to other baselines, SmartFL surpasses ChatGPT-4 (Log) in the largest number of instances.

We manually analyse the 12 cases where ChatGPT-4 (Log) is missing compared to SmartFL.
Seven of these cases can be attributed to missing logic, examples of which will be illustrated below.

\definecolor{diffstart}{RGB}{0,0,0}
\definecolor{diffincl}{RGB}{23, 117, 91}
\definecolor{diffrem}{RGB}{255, 76, 11}

\lstdefinelanguage{diff}{
basicstyle=\ttfamily,
morecomment=[f][\color{diffstart}]{@@},
morecomment=[f][\color{diffincl}]{+\ },
morecomment=[f][\color{diffrem}]{-\ },
}

\newmdenv[
  leftline=false,
  rightline=false,
  topline=false,
  bottomline=false,
  backgroundcolor=texgray, 
  linewidth=2pt,
  skipabove=5pt
]{mymiddleline}

\begin{mymiddleline}
\tiny
\begin{lstlisting}[
  language=diff,
  basicstyle=\footnotesize\ttfamily,
  breaklines=true,
  lineskip=-10pt, %
]
if (pfxLen > 0) {
+   char firstSigDigit = 0;
+   for(int i = pfxLen; i < str.length(); i++) {
+       firstSigDigit = str.charAt(i);
+       if (firstSigDigit == '0') {
+           pfxLen++;
+       } else {
+           break;
+       }
+   }
    final int hexDigits = str.length() - pfxLen;
-   if (hexDigits > 16) {
+   if (hexDigits > 16 || (hexDigits==16 && firstSigDigit > '7')) {
        return createBigInteger(str);
    }
-   if (hexDigits > 8) {
+   if (hexDigits > 8 || (hexDigits==8 && firstSigDigit > '7')) {
        return createLong(str);
    }
    return createInteger(str);
}
\end{lstlisting}
\end{mymiddleline}

\vspace{0.2cm}
The above code snippet shows that addressing such faults requires adding more logical code, without modifying the original programs.
This implies that ChatGPT-4 (Log) exhibits insufficient performance on localising faults associated with logical inadequacies.
To improve the effectiveness of ChatGPT-4 (Log) in fault localisation, future enhancements could focus on augmenting its comprehension and identification of such missing logical code.
\vspace{0.2cm}
\begin{mdframed}[backgroundcolor=texgray, linewidth=0.5pt]
\textbf{Answer to RQ1.2:}
ChatGPT-4 (Log) can identify numerous unique faults that are untraceable by alternative methods.
For example, compared to SmartFL (the state-of-the-art baseline), there are on average 44 (23.66\%) faults that ChatGPT-4 (Log) can localise but SmartFL cannot.
\end{mdframed}

\noindent
\textbf{RQ1.3: Consistency of ChatGPT:}
In this section, we assess the consistency of ChatGPT's fault localisation, focusing on variability across repeated experiments.
We present Table~\ref{Single Results of Repeated Experiments} showcasing results on the Defects4J dataset from five repeated experiments.

\begin{table}[htp]

\scriptsize
\caption{RQ1.3: Results of Repeated Experiments on the Defects4J Dataset}
\label{Single Results of Repeated Experiments}
\centering
\renewcommand\arraystretch{1}
\resizebox{\linewidth}{!}{
\begin{tabular}{cc|ccccc||c|ccccc}
\hline 
\multirow{2}{*}{Method} & \multirow{2}{*}{Time} & \multicolumn{5}{c||}{$TOP$-$N$} & \multirow{2}{*}{Method} & \multicolumn{5}{c}{$TOP$-$N$}\tabularnewline
 &  & $TOP$-$1$ & $TOP$-$2$ & $TOP$-$3$ & $TOP$-$4$ & $TOP$-$5$ &  & $TOP$-$1$ & $TOP$-$2$ & $TOP$-$3$ & $TOP$-$4$ & $TOP$-$5$\tabularnewline
\hline 
\multirow{5}{*}{\makecell[c]{ChatGPT-3.5\\(Origin)}} & 1 & 80 & 123 & 143 & 153 & 164 & \multirow{5}{*}{\makecell[c]{ChatGPT-4\\(Origin)}} & 100 & 139 & 157 & 172 & 180\tabularnewline
 & 2 & 82 & 119 & 136 & 151 & 158 &  & 94 & 125 & 144 & 159 & 171\tabularnewline
 & 3 & 81 & 114 & 134 & 146 & 159 &  & 99 & 133 & 155 & 163 & 172\tabularnewline
 & 4 & 88 & 125 & 137 & 151 & 162 &  & 93 & 126 & 156 & 169 & 180\tabularnewline
 & 5 & 77 & 112 & 130 & 142 & 154 &  & 92 & 125 & 150 & 165 & 178\tabularnewline
\hline 
\multicolumn{2}{c|}{\textbf{Average}} & 81.6 & 118.6 & 136.0 & 148.6 & 159.4 & \multicolumn{1}{c|}{\textbf{Average}} & 95.6 & 129.6 & 152.4 & 165.6 & 176.2\tabularnewline
\hline 
\hline 
\multirow{5}{*}{\makecell[c]{ChatGPT-3.5\\(Log)}} & 1 & 89 & 131 & 149 & 156 & 167 & \multirow{5}{*}{\makecell[c]{ChatGPT-4\\(Log)}} & 135 & 170 & 187 & 198 & 208\tabularnewline
 & 2 & 87 & 132 & 149 & 162 & 165 &  & 136 & 170 & 189 & 197 & 207\tabularnewline
 & 3 & 86 & 118 & 141 & 155 & 164 &  & 145 & 174 & 190 & 199 & 207\tabularnewline
 & 4 & 93 & 126 & 146 & 163 & 175 &  & 142 & 173 & 193 & 208 & 217\tabularnewline
 & 5 & 85 & 125 & 145 & 154 & 165 &  & 136 & 172 & 188 & 200 & 209\tabularnewline
\hline 
\multicolumn{2}{c|}{\textbf{Average}} & 88.0 & 126.4 & 146.0 & 158.0 & 167.2 & \multicolumn{1}{c|}{\textbf{Average}} & 138.8 & 171.8 & 189.4 & 200.4 & 209.6\tabularnewline
\hline 
\end{tabular}
}
\end{table}

As illustrated in Table~\ref{Single Results of Repeated Experiments}, despite minor fluctuations attributable to randomness, a consistent pattern emerges regarding the comparative performance.
ChatGPT-4 (Log) consistently outperforms the competition, ranking as the top performer in all five repetitions.
This evidence serves to reinforce our conclusions from RQ1.1.

To further assess consistency, we present Table~\ref{Overall Results of Repeated Experiments} summarising the variances observed in repeated experiments.
The variance acts as a consistency metric, with lower values indicating higher consistency and insights into each method's reliability.

\begin{table}[htp]
\scriptsize
\caption{RQ1.3: Variance Results of Repeated Experiments}
\label{Overall Results of Repeated Experiments}
\centering
\renewcommand\arraystretch{1}
\resizebox{\linewidth}{!}{
\begin{tabular}{cc|ccccc||c|ccccc}
\hline 
\multirow{2}{*}{Method} & \multirow{2}{*}{Program} & \multicolumn{5}{c||}{Variance of $TOP$-$N$ value} & \multirow{2}{*}{Method} & \multicolumn{5}{c}{Variance of $TOP$-$N$ value}\tabularnewline
 &  & $TOP$-$1$ & $TOP$-$2$ & $TOP$-$3$ & $TOP$-$4$ & $TOP$-$5$ &  & $TOP$-$1$ & $TOP$-$2$ & $TOP$-$3$ & $TOP$-$4$ & $TOP$-$5$\tabularnewline
\hline 
\multirow{6}{*}{\makecell[c]{ChatGPT-3.5\\(Origin)}} & 
Chart & 0.17 & 0.21 & 0.24 & 0.24 & 0.25 & \multirow{6}{*}{\makecell[c]{ChatGPT-4\\(Origin)}} & 0.25 & 0.22 & 0.18 & 0.17 & 0.17\tabularnewline
& Lang & 0.19 & 0.24 & 0.25 & 0.25 & 0.25 &  & 0.22 & 0.24 & 0.25 & 0.25 & 0.25\tabularnewline
& Math & 0.21 & 0.25 & 0.25 & 0.25 & 0.24 &  & 0.22 & 0.25 & 0.24 & 0.22 & 0.21\tabularnewline
& Mockito & 0.22 & 0.14 & 0.12 & 0.10 & 0.09 &  & 0.22 & 0.20 & 0.15 & 0.12 & 0.12\tabularnewline
& Time & 0.13 & 0.20 & 0.22 & 0.24 & 0.25 &  & 0.14 & 0.18 & 0.21 & 0.23 & 0.25\tabularnewline
& Closure & 0.14 & 0.20 & 0.23 & 0.24 & 0.25 &  & 0.16 & 0.21 & 0.24 & 0.25 & 0.25\tabularnewline
\hline 
\multicolumn{2}{c|}{\textbf{Average}} & 0.13 & 0.14 & 0.12 & 0.10 & 0.09 & \multicolumn{1}{c|}
{\textbf{Average}} & \cellcolor[HTML]{fce5cd}\textit{0.14} & \cellcolor[HTML]{fce5cd}\textit{0.18} & \cellcolor[HTML]{fce5cd}\textit{0.15} & \cellcolor[HTML]{fce5cd}\textit{0.12} & \cellcolor[HTML]{fce5cd}\textit{0.12}\tabularnewline
\hline 
\hline 
\multirow{6}{*}{\makecell[c]{ChatGPT-3.5\\(Log)}} & Chart & 0.18 & 0.24 & 0.25 & 0.25 & 0.25 & \multirow{6}{*}{\makecell[c]{ChatGPT-4\\(Log)}} & 0.24 & 0.20 & 0.15 & 0.15 & 0.14\tabularnewline
& Lang & 0.19 & 0.24 & 0.25 & 0.25 & 0.25 &  & 0.25 & 0.24 & 0.22 & 0.21 & 0.20\tabularnewline
& Math & 0.23 & 0.25 & 0.25 & 0.24 & 0.23 &  & 0.25 & 0.22 & 0.20 & 0.17 & 0.15\tabularnewline
& Mockito & 0.20 & 0.13 & 0.10 & 0.09 & 0.08 &  & 0.12 & 0.10 & 0.08 & 0.07 & 0.07\tabularnewline
& Time & 0.17 & 0.23 & 0.25 & 0.25 & 0.24 &  & 0.22 & 0.23 & 0.24 & 0.25 & 0.22\tabularnewline
& Closure & 0.13 & 0.21 & 0.23 & 0.24 & 0.25 &  & 0.23 & 0.25 & 0.25 & 0.25 & 0.24\tabularnewline
\hline 
\multicolumn{2}{c|}{\textbf{Average}} & 0.13 & 0.13 & 0.10 & 0.09 & 0.08 & \multicolumn{1}{c|}{\textbf{Average}} & \cellcolor[HTML]{d9ead3}0.12 & \cellcolor[HTML]{d9ead3}0.10 & \cellcolor[HTML]{d9ead3}0.08 & \cellcolor[HTML]{d9ead3}0.07 & \cellcolor[HTML]{d9ead3}0.07\tabularnewline
\hline 
\end{tabular}}
\end{table}

Table~\ref{Overall Results of Repeated Experiments} reveals that among all ChatGPT-related techniques, ChatGPT-4 (Log) has the lowest variance.
The minimal fluctuation in ChatGPT-4 (Log) signifies it as the most consistent approach, further supporting its reliability for fault localisation.
This indicates that the dynamic information in $Prompt_2$, including test cases and error logs generated from their execution, not only enhances ChatGPT-4's fault localisation but also improves result consistency.

Through consistency assessments, we demonstrate the robustness of our results against the inherent variability present in LLM-based techniques.
Furthermore, we highlight the enhanced consistency exhibited by ChatGPT-4 (Log), further validating its potential as a consistent instrument for fault localisation.
Therefore, in subsequent experiments, we will primarily use ChatGPT-4 (Log) as the representative ChatGPT method for experimental result comparison.
\vspace{0.2cm}
\begin{mdframed}[backgroundcolor=texgray, linewidth=0.5pt]
\textbf{Answer to RQ1.3:}
ChatGPT-4 (Log) consistently exhibits superior performance and low variance across multiple assessments compared with ChatGPT-3.5 (Origin/Log) and ChatGPT-4 (Origin).
ChatGPT-4 (Log) has a mean variance of 0.12, 0.10, 0.08, 0.07 and 0.07 in terms of $TOP$-$1$ to $TOP$-$5$ metrics, which is the smallest among all ChatGPT-related methods.
\end{mdframed}

\subsection{RQ2: Impact of Prompt Design}

To address RQ2, we conduct an ablation study focusing on prompt engineering in ChatGPT-4.
Specifically, by excluding distinct elements from the original prompts and comparing the resulting fault localisation outcomes, we aim to discern the individual contributions of each component to the overall efficacy of the system.

In addition to the basic mandatory requirements on fault localisation, we extract four optional components from $Prompt_1$ and $Prompt_2$.
These include the requirement to sort statements in descending order of suspicion in $Prompt_1$ and $Prompt_2$;
the requirement to explain the reason for fault localisation in $Prompt_1$ and $Prompt_2$;
the information about the error log in $Prompt_2$;
and the information about the test cases in $Prompt_2$, respectively.

By excluding each component individually, we created and tested four variant prompt versions:

\begin{itemize}
\item ChatGPT-4 $_{NoOrder}$: This variant removes the need for ChatGPT to rank the potential faulty statements in descending order of suspicion from $Prompt_1$ and $Prompt_2$.

\item ChatGPT-4 $_{NoExplain}$: This variant removes the need for explaining the reasoning behind the current fault localisation results from $Prompt_1$ and $Prompt_2$.

\item ChatGPT-4 $_{NoError}$: This variant removes the error logs from $Prompt_2$, with $Prompt_1$ remaining unchanged.

\item ChatGPT-4 $_{NoTestCase}$: This variant removes the content of test cases from $Prompt_2$, with $Prompt_1$ remaining unchanged.
\end{itemize}

Table~\ref{results of ablation study} presents a comparative analysis of these four ablation methods with the original ChatGPT-4 (Origin) and ChatGPT-4 (Log) on the Defects4J dataset.

\begin{table}[htp]
\scriptsize
\caption{RQ2: Experimental Results of ChatGPT-4 on the Defects4J Dataset under Different Prompts}
\label{results of ablation study}
\centering
\renewcommand\arraystretch{1}
\begin{tabular}{l|ccccc}
\hline 
\multirow{2}{*}{Method} & \multicolumn{5}{c}{$TOP$-$N$}\tabularnewline
 & $TOP$-$1$ & $TOP$-$2$ & $TOP$-$3$ & $TOP$-$4$ & $TOP$-$5$\tabularnewline
\hline 
ChatGPT-4 (Origin) & 95.6 & 129.6 & 152.4 & 165.6 & 176.2\tabularnewline
ChatGPT-4 (Origin)$_{NoOrder}$ & 87.2 & 134.8 & 157.4 & 168.0 & 177.8\tabularnewline
ChatGPT-4 (Origin)$_{NoExplain}$ & 91.2 & 121.8 & 138.2 & 147.4 & 154.2\tabularnewline
\hline
ChatGPT-4 (Log) & 138.8 & 171.8 & 189.4 & 200.4 & 209.6\tabularnewline
ChatGPT-4 (Log)$_{NoOrder}$ & 105.2 & 155.4 & 184.2 & 196.0 & 204.6\tabularnewline
ChatGPT-4 (Log)$_{NoExplain}$ & 132.6 & 164.6 & 179.8 & 187.6 & 192.8\tabularnewline
ChatGPT-4 (Log)$_{NoError}$ & 103.2 & 141.4 & 165.4 & 174.8 & 184.6\tabularnewline
ChatGPT-4 (Log)$_{NoTestCase}$ & 136.8 & 172.4 & 193.0 & 203.4 & 211.4\tabularnewline
\hline 
\end{tabular}
\end{table}

The results in Table~\ref{results of ablation study} reveal several insightful observations:

\begin{itemize}

\item Both the original ChatGPT-4 (Origin) and ChatGPT-4 (Log) achieve the highest accuracy under $TOP$-$1$ metric.

\item The most detrimental exclusion is the error logs in ChatGPT-4 (Log)$_{NoError}$, which substantially diminishes the accuracy across evaluation metrics. This implies that the error log provides a vital dynamic context for ChatGPT-4 to effectively localise faults.

\item ChatGPT-4 (Origin)$_{NoExplain}$ and ChatGPT-4 (Log)$_{NoExplain}$, which lack the requirement to rank potential faults, exhibit noticeable performance degradation in terms of $TOP$-$N$ metrics.

\item Removing test case details in ChatGPT-4 (Log)$_{NoTestCase}$ has minimal impact on performance, and ChatGPT-4 (Log)$_{NoTestCase}$ even achieves better performance in some cases.

\item ChatGPT-4 (Origin)$_{NoOrder}$ and ChatGPT-4 (Log)$_{NoOrder}$ demonstrate divergent performance effects. The accuracy of ChatGPT-4 (Log)$_{NoOrder}$ is overall inferior to ChatGPT-4 (Log), whereas relative to ChatGPT-4 (Origin), ChatGPT-4 (Origin)$_{NoOrder}$ exhibits a performance decline only in terms of $TOP$-$1$ metric.

\end{itemize}

Furthermore, we conduct statistical significance analysis using the Wilcoxon signed-rank test, as shown in Table~\ref{p-value}.
In hypothesis testing, the null hypothesis states that the performance of the ChatGPT-4 variants is not inferior to the original ChatGPT-4.
Each value in the table represents the P-value when comparing the method on the y-axis against the method on the x-axis for the corresponding metric.
If the obtained P-value falls below the typical alpha significance level of 0.05, the null hypothesis is consequently rejected, indicating a significant performance difference.

\begin{table}[htp]
\scriptsize
\caption{RQ2: P-values of Ablation Study}
\label{p-value}
\centering
\renewcommand\arraystretch{1}
\begin{tabular}{l|ccccc|ccccc}
\hline 
\multirow{2}{*}{Method} & \multicolumn{5}{c|}{Origin} & \multicolumn{5}{c}{Log}\tabularnewline
 & $TOP$-$1$ & $TOP$-$2$ & $TOP$-$3$ & $TOP$-$4$ & $TOP$-$5$ & $TOP$-$1$ & $TOP$-$2$ & $TOP$-$3$ & $TOP$-$4$ & $TOP$-$5$\tabularnewline
\hline
ChatGPT-4$_{NoOrder}$ &\cellcolor[HTML]{d9ead3} 0.0068 & 0.75 & 0.72 & 0.58 & 0.5 &\cellcolor[HTML]{d9ead3} 4.8e-06 &\cellcolor[HTML]{d9ead3} 9.9e-05 &\cellcolor[HTML]{d9ead3} 0.043 & 0.19 & 0.15\tabularnewline
ChatGPT-4$_{NoExplain}$ &\cellcolor[HTML]{d9ead3} 0.039 &\cellcolor[HTML]{d9ead3} 0.021 &\cellcolor[HTML]{d9ead3} 0.00032 &\cellcolor[HTML]{d9ead3} 4.1e-05 &\cellcolor[HTML]{d9ead3} 3.5e-05 &\cellcolor[HTML]{d9ead3} 0.084 &\cellcolor[HTML]{d9ead3} 0.0093 &\cellcolor[HTML]{d9ead3} 0.0038 &\cellcolor[HTML]{d9ead3} 0.0011 &\cellcolor[HTML]{d9ead3} 8.3e-05\tabularnewline
ChatGPT-4$_{NoError}$ & - & - & - & - & - &\cellcolor[HTML]{d9ead3} 9.3e-10 &\cellcolor[HTML]{d9ead3} 5.3e-06 &\cellcolor[HTML]{d9ead3} 1.3e-05 &\cellcolor[HTML]{d9ead3} 3.7e-06 &\cellcolor[HTML]{d9ead3} 6e-06\tabularnewline
ChatGPT-4$_{NoTestCase}$ & - & - & - & - & - & 0.15 & 0.61 & 0.8 & 0.77 & 0.73\tabularnewline
\hline 
\end{tabular}
\end{table}

The P-values shown in Table ~\ref{p-value} indicate that ChatGPT-4 (Log)$_{NoOrder}$, ChatGPT-4 (Origin)$_{NoExplain}$ and ChatGPT-4 (Log)$_{ NoError}$ perform significantly differently from standard ChatGPT-4 (Origin) or ChatGPT-4 (Log) in some cases.
These results suggest that asking ChatGPT to rank statements, explain their reasoning, or provide ChatGPT with access to error log can improve ChatGPT-4's fault localisation capabilities in specific cases.
\vspace{0.2cm}
\begin{mdframed}[backgroundcolor=texgray, linewidth=0.5pt]
\textbf{Answer to RQ2:}
Each component within the prompts we design makes a positive contribution to the effectiveness of ChatGPT-4.
The most pronounced impact arises from the error log messages, excluding which causes the effectiveness to decline by 25.6\% under the $TOP$-$1$ metric.
\end{mdframed}

\subsection{RQ3: Impact of Context}

In previous experiments, we consistently used the function in the Defects4J dataset that contains the faulty statements as the context in prompts.
This design is based on the premise that the function scope offers an adequate understanding of statement behaviour to characterise the faults.
To further evaluate ChatGPT's fault localisation capability and reliability in practical development scenarios, we examine whether the length of code context provided in the prompt can influence its performance.
Therefore, we conduct two controlled experiments by manipulating the context range within the prompts.
Specifically, we carry out a ``narrowing context experiment'' and an ``expanding context experiment''.
In addition, to ensure the fairness of the comparison, in this RQ, the range of fault location on all baselines is consistent with the adjusted context.

In the narrowing context experiment, we constrain the context to five lines of code before and after the faulty statement within its encapsulating function.
It should be noted that the ideal maximum values for the evaluation metrics in this study differ from those in RQ1.
In RQ1, each prompt only contains one function regardless of how many faulty statements are in the function.
Finally, we construct independent prompts for a total of 528 faulty statements.
The results of the experiment with narrowed context are shown in Table~\ref{Result of limited Context}.

\begin{table}[htp]
 \tiny
\caption{RQ3: Experimental Results on the Defects4J Dataset of Narrowed Context}
\label{Result of limited Context}
\centering
\renewcommand\arraystretch{1}
\begin{tabular}{c|c|cc|cc|c|cc}
\hline 
\multirow{3}{*}{Program} & \multirow{3}{*}{$TOP$-$N$} & \multicolumn{7}{c}{Method}\tabularnewline
 &  & \multicolumn{2}{c|}{SBFL} & \multicolumn{2}{c|}{MBFL} & \multirow{2}{*}{SmartFL} & \multicolumn{2}{c}{ChatGPT-4}\tabularnewline
 &  & $Ochiai$ & $Dstar$ & $Ochiai$ & $Dstar$ &  & Origin & Log\tabularnewline
\hline 
\multirow{5}{*}{Chart} & $TOP$-$1$ & 4 & 4 & 6 & 6 & 11 & 10.4 & 13.6\tabularnewline
 & $TOP$-$2$ & 13 & 13 & 22 & 22 & 13 & 16.6 & 17.0\tabularnewline
 & $TOP$-$3$ & 18 & 18 & 28 & 28 & 14 & 19.2 & 21.8\tabularnewline
 & $TOP$-$4$ & 22 & 22 & 28 & 28 & 15 & 20.0 & 22.0\tabularnewline
 & $TOP$-$5$ & 23 & 23 & 28 & 28 & 17 & 20.2 & 22.2\tabularnewline
\hline 
\multirow{5}{*}{Lang} & $TOP$-$1$ & 31 & 31 & 18 & 14 & 50 & 45.0 & 65.8\tabularnewline
 & $TOP$-$2$ & 50 & 53 & 52 & 50 & 63 & 66.0 & 80.0\tabularnewline
 & $TOP$-$3$ & 69 & 70 & 62 & 60 & 69 & 76.4 & 86.6\tabularnewline
 & $TOP$-$4$ & 86 & 83 & 70 & 68 & 78 & 80.6 & 88.6\tabularnewline
 & $TOP$-$5$ & 92 & 90 & 78 & 76 & 89 & 82.4 & 89.4\tabularnewline
\hline 
\multirow{5}{*}{Math} & $TOP$-$1$ & 59 & 56 & 50 & 50 & 96 & 114.0 & 131.0\tabularnewline
 & $TOP$-$2$ & 115 & 106 & 88 & 88 & 127 & 153.6 & 164.8\tabularnewline
 & $TOP$-$3$ & 146 & 138 & 106 & 106 & 147 & 173.4 & 180.2\tabularnewline
 & $TOP$-$4$ & 171 & 162 & 114 & 114 & 163 & 178.8 & 185.8\tabularnewline
 & $TOP$-$5$ & 186 & 181 & 130 & 130 & 177 & 181.8 & 189.0\tabularnewline
\hline 
\multirow{5}{*}{Mockito} & $TOP$-$1$ & 19 & 19 & 16 & 16 & - & 24.0 & 31.2\tabularnewline
 & $TOP$-$2$ & 35 & 34 & 50 & 48 & - & 31.8 & 36.4\tabularnewline
 & $TOP$-$3$ & 36 & 35 & 52 & 52 & - & 35.2 & 37.8\tabularnewline
 & $TOP$-$4$ & 37 & 36 & 54 & 54 & - & 36.2 & 37.8\tabularnewline
 & $TOP$-$5$ & 37 & 36 & 56 & 56 & - & 36.4 & 37.8\tabularnewline
\hline 
\multirow{5}{*}{Time} & $TOP$-$1$ & 17 & 18 & 10 & 10 & 16 & 16.4 & 25.6\tabularnewline
 & $TOP$-$2$ & 19 & 20 & 12 & 12 & 20 & 24.2 & 30.0\tabularnewline
 & $TOP$-$3$ & 24 & 24 & 18 & 18 & 25 & 30.4 & 33.6\tabularnewline
 & $TOP$-$4$ & 36 & 37 & 22 & 22 & 37 & 34.2 & 37.4\tabularnewline
 & $TOP$-$5$ & 41 & 41 & 34 & 34 & 39 & 35.4 & 38.2\tabularnewline
\hline 
\multirow{5}{*}{Closure} & $TOP$-$1$ & 65 & 62 & 2 & 2 & - & 58.4 & 72.2\tabularnewline
 & $TOP$-$2$ & 79 & 74 & 12 & 12 & - & 81.8 & 90.6\tabularnewline
 & $TOP$-$3$ & 88 & 82 & 20 & 20 & - & 89.4 & 98.8\tabularnewline
 & $TOP$-$4$ & 98 & 93 & 22 & 22 & - & 95.6 & 102.0\tabularnewline
 & $TOP$-$5$ & 104 & 99 & 28 & 28 & - & 97.6 & 103.4\tabularnewline
\hline
\hline
\multirow{5}{*}{\textbf{Average}} & $TOP$-$1$ & 32.50 & 31.67 & 17.00 & 16.33 & 43.25 & 44.70 & \cellcolor[HTML]{d9ead3}56.57\tabularnewline
 & $TOP$-$2$ & 51.83 & 50.00 & 39.33 & 38.67 & 55.75 & 62.33 & \cellcolor[HTML]{d9ead3}69.80\tabularnewline
 & $TOP$-$3$ & 63.50 & 61.17 & 47.67 & 47.33 & 63.75 & 70.67 & \cellcolor[HTML]{d9ead3}76.47\tabularnewline
 & $TOP$-$4$ & 75.00 & 72.17 & 51.67 & 51.33 & 73.25 & 74.23 &\cellcolor[HTML]{d9ead3} 78.93\tabularnewline
 & $TOP$-$5$ &\cellcolor[HTML]{d9ead3} 80.50 & 78.33 & 59.00 & 58.67 &\cellcolor[HTML]{d9ead3} 80.50 & 75.63 & 80.00\tabularnewline
\hline 
\end{tabular}
\end{table}

From Table~\ref{Result of limited Context}, we find ChatGPT-4 (Log) generally maintains advantages.
Specifically, on average, ChatGPT-4 (Log) achieves the best results under $TOP$-$1$ to $TOP$-$4$ metrics.
However, with the reduction of code context, the performance of other baselines has also improved, making ChatGPT-4 (Log) less leading, and even surpassed by SBFL and SmartFL in terms of $TOP$-$5$ metric.

\begin{table}[htp]
 \scriptsize
\caption{RQ3: Experimental Results on the Defects4J Dataset of Expanding Context}
\label{Result of Expanding Context}
\centering
\renewcommand\arraystretch{1}
\begin{tabular}{c|c|cc|cc|c|cc}
\hline 
\multirow{3}{*}{Program} & \multirow{3}{*}{$TOP$-$N$} & \multicolumn{7}{c}{Method}\tabularnewline
 &  & \multicolumn{2}{c|}{SBFL} & \multicolumn{2}{c|}{MBFL} & \multirow{2}{*}{SmartFL} & \multicolumn{2}{c}{ChatGPT-4}\tabularnewline
 &  & $Ochiai$ & $Dstar$ & $Ochiai$ & $Dstar$ &  & Origin & Log\tabularnewline
\hline 
\multirow{5}{*}{Chart} & $TOP$-$1$ & 0 & 0 & 0 & 0 & 3 & 1.0 & 1.8\tabularnewline
 & $TOP$-$2$ & 1 & 0 & 1 & 1 & 4 & 1.8 & 1.8\tabularnewline
 & $TOP$-$3$ & 3 & 1 & 1 & 1 & 4 & 1.8 & 1.8\tabularnewline
 & $TOP$-$4$ & 3 & 1 & 1 & 1 & 4 & 1.8 & 1.8\tabularnewline
 & $TOP$-$5$ & 3 & 1 & 1 & 1 & 4 & 1.8 & 1.8\tabularnewline
\hline 
\multirow{5}{*}{Lang} & $TOP$-$1$ & 0 & 0 & 0 & 0 & 3 & 1.0 & 2.6\tabularnewline
 & $TOP$-$2$ & 0 & 0 & 3 & 3 & 3 & 1.4 & 3.4\tabularnewline
 & $TOP$-$3$ & 0 & 0 & 4 & 4 & 4 & 1.8 & 3.6\tabularnewline
 & $TOP$-$4$ & 1 & 1 & 4 & 4 & 4 & 2.0 & 3.6\tabularnewline
 & $TOP$-$5$ & 1 & 1 & 4 & 4 & 4 & 2.0 & 3.6\tabularnewline
\hline 
\multirow{5}{*}{Math} & $TOP$-$1$ & 5 & 4 & 6 & 6 & 5 & 0.8 & 2.4\tabularnewline
 & $TOP$-$2$ & 7 & 6 & 9 & 9 & 9 & 2.2 & 5.4\tabularnewline
 & $TOP$-$3$ & 9 & 6 & 15 & 15 & 11 & 4.0 & 6.2\tabularnewline
 & $TOP$-$4$ & 10 & 6 & 16 & 16 & 13 & 5.4 & 6.4\tabularnewline
 & $TOP$-$5$ & 10 & 6 & 17 & 17 & 14 & 5.6 & 6.8\tabularnewline
\hline 
\multirow{5}{*}{Mockito} & $TOP$-$1$ & 1 & 1 & 0 & 0 & - & 0.2 & 2.4\tabularnewline
 & $TOP$-$2$ & 2 & 1 & 1 & 0 & - & 2.0 & 4.6\tabularnewline
 & $TOP$-$3$ & 2 & 1 & 1 & 1 & - & 2.6 & 5.0\tabularnewline
 & $TOP$-$4$ & 2 & 1 & 1 & 1 & - & 2.6 & 5.0\tabularnewline
 & $TOP$-$5$ & 2 & 1 & 2 & 2 & - & 2.6 & 5.0\tabularnewline
\hline 
\multirow{5}{*}{Closure} & $TOP$-$1$ & 3 & 2 & 0 & 0 & - & 0.4 & 0.0\tabularnewline
 & $TOP$-$2$ & 3 & 2 & 0 & 0 & - & 1.2 & 0.2\tabularnewline
 & $TOP$-$3$ & 3 & 3 & 0 & 0 & - & 1.4 & 0.4\tabularnewline
 & $TOP$-$4$ & 3 & 3 & 0 & 0 & - & 1.6 & 0.4\tabularnewline
 & $TOP$-$5$ & 5 & 5 & 0 & 0 & - & 2.0 & 0.4\tabularnewline
\hline 
\hline 
\multirow{5}{*}{\textbf{Average}} & $TOP$-$1$ & 1.80  & 1.40  & 1.20  & 1.20  & \cellcolor[HTML]{d9ead3}3.67  & 0.68  & 1.84 \tabularnewline
 & $TOP$-$2$ & 2.60  & 1.80  & 2.80  & 2.60  & \cellcolor[HTML]{d9ead3}5.33  & 1.72  & 3.08 \tabularnewline
 & $TOP$-$3$ & 3.40  & 2.20  & 4.20  & 4.20  & \cellcolor[HTML]{d9ead3}6.33  & 2.32  & 3.40 \tabularnewline
 & $TOP$-$4$ & 3.80  & 2.40  & 4.40  & 4.40  & \cellcolor[HTML]{d9ead3}7.00  & 2.68  & 3.44 \tabularnewline
 & $TOP$-$5$ & 4.20  & 2.80  & 4.80  & 4.80  & \cellcolor[HTML]{d9ead3}7.33  & 2.80  & 3.52 \tabularnewline
\hline 
\end{tabular}
\end{table}

In the subsequent expanding context experiment, we expand the context to encompass the class containing the faulty code.
If a class includes multiple faulty statements across different functions, the code for the whole class is used as context within a single prompt.
The experiment results with expanding context are shown in Table~\ref{Result of Expanding Context}.
Due to ChatGPT's token limitations, overly long classes are excluded; ultimately, the experiment uses 56 classes from five different programs.

Unexpectedly, as shown in Table~\ref{Result of Expanding Context}, ChatGPT-4 no longer holds an advantage.
It exhibits inferior performance compared to baselines across most programs and evaluation metrics under class-level context.
Therefore, when expanding the code context in prompts from function-level to class-level, the fault localisation capability of ChatGPT-4 drastically deteriorates.
\vspace{0.2cm}
\begin{mdframed}[backgroundcolor=texgray, linewidth=0.5pt]
\textbf{Answer to RQ3:}
ChatGPT's fault localisation capability is highly sensitive to the code context scope in the prompts.
With functional-level context, ChatGPT-4 (Log) outperforms all the baseline methods, with 30.8\% higher $TOP$-$1$ on average than state-of-the-art;
When expanding to class-level context, ChatGPT-4's localisation ability decreases significantly, leading to a 49.9\% decline in the $TOP$-$1$ metric with ChatGPT-4 (Log).
\end{mdframed}


\section{Extended Analysis on the Threat of Overfitting}
\label{Extended Analysis}

While Defects4J has been widely used in the fault localisation domain~\cite{Grace2021,zou2019empirical,zakari2020multiple}, there is a possibility that it may have been incorporated into the training set of ChatGPT.
This could confer an unfair advantage to ChatGPT when assessing fault localisation effectiveness based on the Defects4J dataset and cause overfitting.

To address this threat and validate the reliability of our previous observations, we build a more recent dataset, which we refer to as the \textbf{Stu}dent-Generated \textbf{Defects} (\textbf{StuDefects}).
StuDefects contains code generated exclusively after 2022.
This guarantees that it does not appear in the training data of ChatGPT because
ChatGPT's training data only extends up to 2021.
To facilitate reproducibility and future research, we have open-sourced this dataset on GitHub~\cite{thisweb}.

In particular, StuDefects comes from a university's course submission platform that tracks students' code submissions during classroom programming tasks.
The platform offers task descriptions and uses pre-prepared test cases to compile and run students' code.
It then compares the tests' outputs to expected results to evaluate code correctness.
If any test fails, students are prompted to revise and resubmit their code in an iterative process until they achieve the expected outcome.

During the data collection process, we adhere to the subsequent criteria:

\noindent \textbf{Time Constraint} We include only the code submitted from 2022 onward to guarantee that the code in this dataset is more current than the cut-off date of ChatGPT's training data.\\
\noindent \textbf{Language Specificity} We choose only code in Java language, in alignment with Defects4J, to maintain the consistency of our study in terms of programming languages.\\
\noindent \textbf{Precise Fault Label} We focus on instances where a student submits a faulty code and then rectifies it with a single line modification before resubmitting. This approach allows us to identify the actual location of the fault (serving as our fault localisation label) and ensures that the faults are real instead of artificially injected.\\
\noindent \textbf{Reduced Redundancy} For each programming task, we keep only one faulty code, thus removing the redundancy of code in our dataset.


Following the criteria outlined above while traversing the entire database, we successfully identify 77 
unique programs from different programming tasks.
These tasks encompass a variety of algorithmic complexities, from simple tasks such as basic mathematical operations and string manipulations, to more challenging ones like dynamic programming and large number handling.
We also pair each faulty program with its corresponding rectified version.
The length of these programs ranges from 5 to 86 lines, with an average of 25.7 lines.

\begin{wraptable}{l}{7cm}
\scriptsize
\caption{Extended Analysis: Fault Localisation Results on the StuDefects in Terms of $TOP$-$N$ Metric }
\label{generalresultsNew}
\begin{tabular}{c|cc|cc|c}
\hline 
\multirow{3}{*}{$TOP$-$N$} & \multicolumn{5}{c}{Method}\tabularnewline
 & \multicolumn{2}{c|}{SBFL} & \multicolumn{2}{c|}{MBFL} & \multirow{2}{*}{\makecell[c]{ChatGPT-4\\(Origin)}}\tabularnewline
 & $Ochiai$ & $Dstar$ & $Ochiai$ & $Dstar$ & \tabularnewline
\hline 
$TOP$-$1$ & 3 & 3 & 17 & 17 & \cellcolor[HTML]{d9ead3}26.0\tabularnewline
$TOP$-$2$ & 11 & 11 & 25 & 25 & \cellcolor[HTML]{d9ead3}36.2\tabularnewline
$TOP$-$3$ & 15 & 15 & 33 & 34 & \cellcolor[HTML]{d9ead3}41.4\tabularnewline
$TOP$-$4$ & 18 & 18 & 42 & 42 & \cellcolor[HTML]{d9ead3}44.2\tabularnewline
$TOP$-$5$ & 24 & 24 & \cellcolor[HTML]{d9ead3}48 & \cellcolor[HTML]{d9ead3}48 & 46.0\tabularnewline
\hline 
\end{tabular}
\end{wraptable}

Table~\ref{generalresultsNew} shows the experimental results based on the StuDefects dataset.
Given StuDefects’ smaller scale and being within ChatGPT's token limit, the entire code (all with functional level) is used as the context. 
We focus on ChatGPT-4 (Origin) in this extended analysis because the failed test case log information of StuDefects is different from that of Defects4J.


As discerned from Table~\ref{generalresultsNew}, ChatGPT-4 (Origin) demonstrates superiority in most scenarios.
Furthermore, ChatGPT-4 (Origin) surpasses the second-best MBFL by 52.9\% in $TOP$-$1$ evaluation metric.
These experimental findings indicate that ChatGPT exhibits consistently promising fault localisation performance on both Defects4J and the dataset beyond its training data cut-off in 2021.

\section{Related Work}
\label{RelatedWork}

\subsection{Fault Localisation}

To assist developers for localising faults, various approaches have been proposed.
These approaches include SBFL~\cite{abreu2007accuracy}, MBFL~\cite{moon2014ask}, and deep learning-based fault localisation~\cite{DeepFL2019}.

Among traditional fault localisation techniques, SBFL stands out for its simplicity and low computational cost, attracting significant attention.
However, it faces challenges such as inaccuracy and the issue of assigning identical suspiciousness scores to multiple statements~\cite{feyzi2018program, steimann2013threats,7390282}.
Similarly, while MBFL is acknowledged for its accuracy, it is hampered by high computational costs and difficulties in modeling complex software faults.
This limits its applicability to large-scale programs and affects its effectiveness in localising complex faults in industrial programs such as Defects4J~\cite{li2020hmer, steimann2013threats,7390282}.

To address these limitations, recent studies have aimed to improve localisation accuracy by incorporating additional semantic information.
For example, Zeng et al.~\cite{zeng2022fault} proposed using probabilistic models or machine learning to capture program semantics. They introduced SmartFL, a probabilistic model-based technique that considers only value correctness, not full semantics.
Using static and dynamic analysis, they constructed a probabilistic model and ranked elements by maximum posterior probability.
Their method achieved 21\% $TOP$-$1$ statement-level accuracy on Defects4J, outperforming other baselines.

With the extensive application of neural networks, the use of deep learning in software testing has motivated many studies on fault localisation, such as Grace~\cite{Grace2021}, FLUCCS~\cite{sohn2017fluccs}, and DeepFL~\cite{DeepFL2019}.

Sohn et al.~\cite{sohn2017fluccs} proposed FLUCCS, which leverages code and change metrics to improve the ranking of likely faulty program elements.
FLUCCS expands SBFL by incorporating additional source code features, including size, age, and code churn.
Through genetic programming and linear rank SVM, FLUCCS integrates suspiciousness values from SBFL with code metrics into a unified ranking function.
Evaluation of 210 faults in Defects4J showed FLUCCS significantly outperformed SBFL, ranking the faulty method first for over half the faults.
FLUCCS demonstrates code and change metrics can enhance the accuracy and effort reduction of fault localisation.

Li et al.~\cite{DeepFL2019} proposed DeepFL, integrating suspiciousness-based, fault-proneness-based, and text-similarity-based features from different domains.
DeepFL significantly outperformed existing learning-based techniques on Defects4J, and they studied the impact of model configurations and features.
DeepFL also evidenced high effectiveness for cross-project prediction.

Subsequently, Lou et al.~\cite{Grace2021} proposed a graph-based representation learning technique Grace to boost coverage-based fault localisation.
They constructed a graph linking test cases and program entities based on coverage relationship and code structure.
A gated graph neural network learned valuable features from the graph representation. This technique substantially exceeded existing fault localisation techniques on Defects4J.
Precisely, Grace could localise 195 bugs in terms of $TOP$-$1$ metric, whereas the best compared technique, DeepFL, could at most localise 166 bugs.

While deep learning-based fault localisation methods exhibit promising development potential, including the aforementioned, they typically localise only to the faulty function rather than the faulty statement.
Therefore, we do not select such techniques as baselines in this paper.

\subsection{Large Language Model for Software Engineering}

Given the breakthrough of large language models (LLMs), studies have sought to explore how LLMs could assist developers in a variety of tasks, including:

Code generation:
LLMs can generate standard-compliant code based on specific requirements or conditions.
For example,
Dong et al.~\cite{dong2023self} explored a self-collaboration framework employing LLMs, specifically ChatGPT, for code generation.
By simulating different roles such as analysts, coders, and testers within a virtual team, their approach achieved state-of-the-art performance and even surpassed GPT-4 in complex code generation tasks.
Their framework offers a new direction for leveraging the capabilities of LLMs to handle real-world complex programming problems efficiently.

Program repair: 
LLMs can be used to generate prompts that guide developers to improve their code quality, perform refactoring operations, or fix the errors or bugs in their code.
For example, 
Dominik et al.~\cite{sobania2023analysis} evaluated ChatGPT's bug-fixing capabilities on the QuixBugs benchmark, comparing it to deep learning methods like CoCoNut and Codex.
The study finds that ChatGPT, enhanced by its dialogue system for additional context, fixes 31 out of 40 bugs, outperforming state-of-the-art approaches.
Wei et al.~\cite{fse2023copilot} introduced Rectify, a framework that synergistically combines Large Language Models (LLMs) with completion engines for automated program repair.
Their study indicates that Rectify achieves state-of-the-art results on the Defects4J dataset, producing more valid and correct patches within the same generation budget.
Their evaluation on a subset of the widely-used Defects4J 1.2 and 2.0 datasets shows that Rectify fixes 66 and 50 bugs, respectively, surpassing the best-performing baseline by 14 and 16 bugs fixed.
Xie et al.~\cite{xie2023impact} evaluated ten Code Language Models (CLMs) on four Automated Program Repair (APR) benchmarks.
Their study shows that even without fine-tuning, the best CLM fixes 72\% more bugs than state-of-the-art deep learning-based APR techniques.
Fine-tuning further improves CLMs' fixing capabilities, enabling them to fix 31\%-1,267\% more bugs and outperform existing techniques by 46\%-164\%.


Code summarisation:
LLMs can produce summaries explaining the functionality and intent of code segments.
For example,
Sun et al.~\cite{sun2023automatic} used prompts to request ChatGPT to generate comments for all code snippets in the CSN-Python test set.
They compared ChatGPT's automatic code summarisation capabilities with three dedicated code summarisation models (specifically, NCS, CodeBERT, and CodeT5) on the CSN-Python dataset.
Their findings revealed that on the whole, ChatGPT underperformed compared to specialised code summarisation models on the CSN-Python dataset in terms of BLEU, METEOR, and ROUGE-L scores.

Test generation:
LLMs can aid in creating diverse test cases, ensuring the robustness and correctness of software applications.
For example,
Liu et al.~\cite{liu2023your} questioned the functional correctness of code generated by LLMs, introducing a rigorous evaluation framework named EvalPlus.
Their extensive evaluation across 19 popular LLMs demonstrates that their framework is capable of catching significant amounts of previously undetected wrong code, shedding light on the need for more robust testing methods in code synthesis.

There has been limited work on LLMs for fault localisation. 
Cao et al.~\cite{cao2023study} included a small-scale test of ChatGPT's fault localisation abilities within their comprehensive assessment of the model's capabilities.
Kang et al.~\cite{kang2023preliminary} proposed AutoFL, an LLM-based fault localisation method.
AutoFL determines the fault location through a two-stage process: first generating a root cause explanation, then predicting the location based on available information.
In a series of experiments, AutoFL exhibited remarkable potential, outperforming traditional techniques on Defects4J dataset.

Existing preliminary studies have demonstrated potential, yet several gaps remain.
Firstly, the advanced ChatGPT-4 has not been used, potentially overlooking its capabilities in fault localisation.
Secondly, a lack of comparison with state-of-the-art fault localisation methods leaves the efficacy of the proposed methods unverified.
Lastly, the effect of different code context lengths on fault localisation has not been studied.
Addressing these gaps is helpful for a comprehensive understanding of LLMs in fault localisation.


\section{Discussion}
\label{Discussion}

\subsection{Discussion on Threats to Validity}

Firstly, ChatGPT's responses contain inherent randomness, potentially affecting our results' credibility.
To mitigate this threat, we perform each experiment five times, striving to guarantee the consistency and robustness of our outcomes.
Therefore, this threat is limited.

Secondly, the Defects4J dataset, widely used in fault localisation, might be a part of ChatGPT's training data, posing a potential issue of data leakage.
However, our experimental findings show that ChatGPT's performance falls behind the baseline when presented with substantial code context. 
This suggests that ChatGPT doesn't simply memorise instances from Defects4J, which alleviates some concerns about data leakage.
In additional experiments with the newly collected dataset, StuDefects, we observe that ChatGPT outperforms the baseline in fault localisation, further reducing concerns about data leakage.

\subsection{Implications for Researchers and Developers}
Based on the results and comparative analysis in this study, we summarise several key implications for researchers and developers seeking to apply large language models like ChatGPT-4 for fault localisation:

\subsubsection{For Researchers:}

\begin{itemize}
\item \textbf{Prompt and Context Optimization:}
Research into optimizing prompt design and context length is crucial, as these elements significantly impact the efficacy of LLMs in fault localisation. Experimentation with different prompt structures and context lengths can lead to enhanced performance and usability.
\item  \textbf{Model Enhancement and Fine-Tuning:}
Despite their capabilities, current LLMs like ChatGPT-4 show limitations in different contexts, necessitating further research to improve their adaptability and accuracy in various settings, especially when the code context expands.
\end{itemize}

\subsubsection{For Developers:}

\begin{itemize}
\item \textbf{Enhanced Debugging:}
The demonstrated superiority of ChatGPT-4 in fault localisation within function-level context can assist developers in more efficiently identifying and addressing software bugs, thereby reducing debugging time.
\item \textbf{Tool Selection:}
Developers need to be discerning in choosing fault localisation methods and should consider the context in which they are working, as LLMs like ChatGPT-4 may not be the best fit in all scenarios, such as when dealing with class-level contexts.
\item \textbf{Error Log Integration:}
Incorporating additional error logs can significantly enhance localisation accuracy and consistency, aiding developers in more effectively leveraging LLMs for fault localisation.
\item \textbf{Awareness of Limitations:}
While LLMs offer considerable benefits in fault localisation, developers should be cognizant of their limitations and employ them judiciously, complementing them with other methods where necessary.
\end{itemize}





\section{Conclusion}
\label{conclusion}

In conclusion, this paper systematically explores the potential of applying ChatGPT to fault localisation in large-scale open-source programs.

When compared to existing fault localisation methods, ChatGPT-4 can outperform them in most situations when the code context of the prompts is within the function-level.
Furthermore, the error log generated during test execution provides substantial assistance for ChatGPT's fault localisation, enhancing performance and consistency.
Moreover, in overlap analysis conducted among all baselines, ChatGPT-4 (Log) consistently exhibits superior performance, showing a higher number of unique instances compared to missing instances.

However, ChatGPT's fault localisation performance is highly sensitive to the length of code context provided.
When the context of the Defects4J dataset expands to the class-level, its performance significantly deteriorates, falling behind that of SmartFL.

Therefore, employing ChatGPT for fault localisation persists as an area requiring further improvements in research, requiring further refinement and investigation to truly leverage its potential advantages in practical scenarios.


\bibliographystyle{ACM-Reference-Format}
\bibliography{sample-base}

\end{document}